\documentstyle[12pt]{article}

\font\fiverm=cmr10

\title{{\bf
Deformation of Schild String 
}}

\author{
{\sc R.Kuriki}$^\dagger$, {\sc S.Ogushi}$^{\dagger\dagger}$
and {\sc A.Sugamoto}$^\clubsuit$\\
\small{\fiverm{Ochanomizu University, Department of Physics, 
Faculty of Science}}\\
\small{\fiverm{1-1 Otsuka 2, Bunkyo-Ku}}
\small{\fiverm{Tokyo}}
\small{\fiverm{112-0012 JAPAN}}\\
\llap{$^\dagger$}%
\small{r-kuriki@cc.ocha.ac.jp}, JSPS Research Fellow\\
\llap{$^{\dagger\dagger}$}%
\small{g9970503@edu.cc.ocha.ac.jp}, JSPS Research Fellow\\
\llap{$^\clubsuit$}%
\small{sugamoto@phys.ocha.ac.jp}\\
}
\date{}

\begin{document}
\baselineskip 0.8cm

\maketitle
\vskip 5mm

{\bf corresponding author}

Rie Kuriki

Ochanomizu University, Department of Physics, Faculty of Science

1-1 Otsuka 2, Bunkyo-Ku Tokyo, 112-0012 JAPAN

TEL : 03 5978 5323 

FAX : 03 5978 5898

e-mail : r-kuriki@cc.ocha.ac.jp

\vskip 5mm

\thispagestyle{empty}
\begin{abstract}

We attempt to construct new superstring actions with a $D$-plet of Majorana fermions 
$\psi^{\cal B}_A$, where ${\cal B}$ is the $D$ dimensional space-time index 
and $A$ is the two dimensional spinor index, by deforming the Schild action.
As a result, we propose three kinds of actions: the first is invariant under $N=1$ (the world-sheet) supersymmetry 
transformation and the area-preserving diffeomorphism. The second contains the Yukawa type interaction.
The last possesses some non-locality because of bilinear terms of $\psi^{\cal B}_A$.
The reasons  
why completing a Schild type superstring action with $\psi^{\cal B}_A$
is difficult are finally discussed.

\end{abstract}

{\centerline {{\bf Keywords}\\
Schild string, $N=1$ two dimensional superspace, deformation}}

\vfill
\vbox{
\hfill  April 1999 \null\par
\hfill OCHA-PP-126}
\null
\clearpage


\section{Introduction}
\indent

Recently the space-time uncertainty relation 
which concerns possible matrix representations of spatial and temporal co-ordinates
is one of the highlighting topics in string theory. 
The non-commutativity of co-ordinates gives rise to a distrust of
the canonical quantization 
in which only a variable for time 
carries an intrinsic meaning, as compared with the other variables for space. 
In other words, when one applies the canonical formulation to a theory 
in which the Lorentz covariance is an important criterion, 
at least one feels inconvenience of the quantization method based on the time-energy uncertainty principle 
because the covariance is not manifest in the Hamiltonian formulation.

There are some proposals for the development of new quantization methods
beyond the canonical one from the motivation above mentioned.
Eguchi 
suggested a novel method to treat the dual-string theory and derived a diffusion equation
where the area, $d\tau{d}\sigma$, parametrizing the world sheet, was regarded as the evolution parameter
\cite{Eguchi}.
Nambu also considered a generalization of the Hamiltonian formulation 
in which both spatial and temporal variables
were dealt with on the same stage \cite{Nambu1}\cite{Nambu2}. 
In his discussion, the following bracket plays an essential role:
\begin{eqnarray}
\{~X^{\cal A}~,~X^{\cal B}~\}~=~\epsilon^{mn}\partial_m{X^{\cal A}}\partial_n{X^{\cal B}},
\label{NB}
\end{eqnarray}
where $m,n$ represents two dimensional vector index, 
$\epsilon^{mn}$ is a two dimensional antisymmetric tensor and 
$X^{\cal A}$ are the bosonic string co-ordinates.
The index ${\cal A}$ denotes the vector index of the embedding space.
One simply confirms that (\ref{NB}) possesses the same properties as the Poisson bracket.
The Schild string \cite{Schild 1} 
described by the square of (\ref{NB}) was used as an available example in \cite{Nambu2}.
In the virtue of (\ref{NB}), Nambu found a link between the Schild string 
and 
the Abelian Yang-Mills theory.
One of the authors (A.S.) tried to apply Nambu's work 
\cite{Nambu2} to the quantization of any other extended objects than strings, such as membranes 
\cite{AS1}. 
Moreover his several works about the application of \cite{Nambu2} 
are summarized in \cite{AS2}.

Our future purpose is to extend the Nambu's work above \cite{Nambu2} to a supersymmetric one. 
We expect that such supersymmetrization will be useful when we will investigate 
super Yang-Mills ($SYM$) theories by applying a kind of Nambu's procedure.
In the view of low energy effective theory for string and in the more fascinating sense between
the $M$-theory and type $IIA$ superstring, the angle of many theoretical physicists points to $SYM$ theories.
For example, 
there is an attractive conjecture suggested by T.Banks, W.Fischler, S.H.Shenker and L.Susskind \cite{BFSS};
an equivalence between the $11$ dimensional $M$-theory and the $U(N)$ $SYM$ theory.
As a candidate for a constructive definition of string theory, a matrix model like the Green-Schwarz 
($GS$) action of type $IIB$ superstring in the Schild gauge is also
proposed in \cite{KKT}.
It is well-known that one of the big and essential problems in elementary particle physics 
is the $QCD$ color confinement in four dimensions.
One needs some dimensional reductions when one applies results from string theory
to understand the $QCD$.
The properities of supersymmetries would be changed via dimensional reductions.
Hence it is significant to construct a new string model which is liked to a $SYM$ theory $\grave{a}~la$ Nambu, 
but would have different supersymmetry from the Schild gauge $GS$ string.
In order to develop the quantum theory of such a new string model in comparison with lower dimensional $QCD$,
for instance supersymmetric $QCD$ in $1+1$ dimensions \cite{MSS},
a super extension of \cite{Nambu2} will be necessary.

In this letter, we will try to build three kinds of Schild type string actions
with a $D$-plet of Majorana fermions 
$\psi^{\cal B}_A$, $(~{\cal B}=0,1,\ldots, D-1)$ transforming in the vector representation of $SO(D-1,1)$.
The capital A denotes world-sheet spinor $SO(1,1)$ indices. Moreover
we make a comparison between one of these actions and two dimensional $U(N)$ $SYM$ action.
Remember that the two dimensional $SYM$ action is not described by ten-dimensional 
$SYM$ like action.
The action in two dimensions 
includes the Yukawa type interaction 
%
$\phi\bar{\Psi}\gamma^5{\Psi}$
%
which is naively required to preserve the supersymmetry
in three dimensions.
Here $\phi$ is a scalar field and $\Psi$ is a Majorana field both in the adjoint representation.
We will choose the supermultiplet of the Neveu-Schwarz-Ramond ($NSR$)
superstring model as our starting point. The multiplet does not contain a
space-time scalar. Maybe we would need some gauge fixing to artificially assemble a field like the adjoint scalar.
From such a reason we recognize that constructing such a kind of Schild type action is not straightforward 
for the $NSR$ superstring.
We will discuss the difficulties in the last section.


\section{Three deformations}
\indent

Firstly let us recall the Schild string \cite{Schild 1} \cite{Yoneya}
\begin{eqnarray}
S_{0}=\int{d^2x}\left[\frac{\Gamma}{e}\{~X^{\cal A}~,~X^{\cal B}~\}^2+\Delta{e}\right],
\label{S_schild}
\end{eqnarray}
where $e$ is an auxiliary field transforming as a scalar density
under two dimensional diffeomorphisms ($Diff_2$)
and $X^{\cal A}$,
${\cal A}=0,1,...3$ are co-ordinates for a string which is propagating in four space-time dimensions.
Here $\Gamma$ and $\Delta$ are arbitrary constants
\footnote{The action (\ref{S_schild}) fixed by $e=1$ and ${\Delta}=0$ corresponds to 
the original string by Schild \cite{Schild 1}:
\begin{eqnarray}
S_{Schild}=\Gamma\int{d^2x}\{X^{\cal A},X^{\cal B}\}^2.
\nonumber
\end{eqnarray}
The gauge symmetry then is the area-preserving diffeomorphisms which are charactrized by 
the following equation:
\begin{eqnarray}
\partial_n\epsilon^n=0,
\nonumber
\end{eqnarray}
where ${\epsilon}^n$ is the infinitesimal parameter of $Diff_2$. 
}. 
The action (\ref{S_schild}) is invariant under $Diff_2$ and is classically equivalent to the Nambu-Goto string
up to the equation of motion for $e$.

\medskip

Supersymmetry can be made obvious by formulating the theory in 
the $N=1$ two dimensional superspace in which the world sheet coordinates $x^m$ is supplemented by
the fermionic coordinates $\theta^\mu$, where $\mu$ is the index for chiralities, $\mu=1,2$.
Let us substitute a scalar superfield $Y^{\cal A}$ for a string co-ordinate $X^{\cal A}$ 
in (\ref{S_schild}):
\begin{eqnarray}
Y^{\cal A}=X^{\cal A}+i\bar{\theta}\psi^{\cal A}+\frac{i}{2}\bar{\theta}\theta{B}^{\cal A},
\end{eqnarray}
where and throughout in the present letter
we use the same notations as the reference \cite{Howe2}
for the representation of two dimensional Clifford algebra.
$\psi^{\cal A}$ is two-component Majorana spinor field and $B^{\cal A}$
is an auxiliary field. 
In the virtue of $B^{\cal A}$ the algebra of supersymmetry transformation is closed off the mass-shell. 
Furthermore we change $d^2x$ to $d^2xd^2\theta$ and fix $e=1,\Delta=0$ in (\ref{S_schild}). 
By integrating super co-ordinates $\theta^\mu$, (\ref{S_schild}) reduces to:
\begin{eqnarray}
S_1&=&\Gamma\int{d^2x}{d^2\theta}\{~Y_{\cal A}~,~Y_{\cal B}~\}^2
\nonumber\\
&=&-2\Gamma\int{d^2x}[
2i\{~X_{\cal A}~,~X_{\cal B}\}\{X^{\cal A},B^{\cal B}\}
+\{~X_{\cal A}~,~X_{\cal B}~\}\{\bar{\psi}^{\cal A},\psi^{\cal B}\}
\nonumber\\
&+&\{\bar{\psi}_{\cal B},X_{\cal A}\}\{\psi^{\cal B},X^{\cal A}\}
-\{\bar{\psi}_{\cal A},X_{\cal B}\}\{\psi^{\cal B},X^{\cal A}\}],
\label{S1}
\end{eqnarray}
which is invariant under the area-preserving diffeomorphism as the result of $e=1$
and the global $N=1$ supersymmetry. Notice that the auxiliary field $B^{\cal A}$ 
becomes a dynamical variable in (\ref{S1}).
From $\delta{S_1}/\delta{B}^{\cal A}=0$, we find 
\begin{eqnarray}
\{\{X^{\cal A},X^{\cal B}\},X_{\cal B}\}=0,
\label{boson}
\end{eqnarray}
which
is the equation of motion for $X^{\cal A}$ derived from the original Schild action.
Hence one can expect that (\ref{S1}) contains a part of the Schild string dynamics.  
However the equation of motion for $X^{\cal A}$ derived from (\ref{S1}) also contains 
the auxiliary field $B^{\cal A}$ :
\begin{eqnarray}
-i\{X_{\cal A},\{X_{\cal C},B^{\cal A}\}\}+i\{X_{\cal A},\{X^{\cal A},B_{\cal C}\}\}
-i\{B^{\cal A},\{X_{\cal C},X_{\cal A}\}\}
\nonumber\\
-\{X_{\cal A},\{\bar{\psi}_{\cal C},\psi^{\cal A}\}\}+\{\bar{\psi}_{\cal A},\{\psi^{\cal A},X_{\cal C}\}\}
-\{\bar{\psi}_{\cal A},\{\psi_{\cal C},X^{\cal A}\}\}=0.
\end{eqnarray}
To solve this equation is not so trivial, it might be possible if we impose a kind of 
canonical commutation relation with respect to the bracket $\{~~,~~\}$ among fields. 
The equation of motion for $\psi^{\cal C}_\mu$ is 
\begin{eqnarray}
\{\{X_{\cal C},X_{\cal A}\},\psi^{\cal A}_{\mu}\}-\{\{X^{\cal A},\psi_{{\cal C}\mu}\},X_{\cal A}\}
+\{\{X_{\cal C},\psi^{\cal A}_\mu\},X_{\cal A}\}=0.
\end{eqnarray}
In this formulation, equations of motion are quite symmetrically expressed among $X^{\cal A}, \psi^{\cal A}$ 
and $B^{\cal A}$.


\bigskip

The $NSR$ superstring is described in $N=1$ superconformal flat 
superspace \cite{Howe2}\cite{Howe1} as:
\begin{eqnarray}
S_{NSR}=\int{d}^2xd^2\theta{E}D_\alpha{Y}^{\cal A}D^\alpha{Y}_{\cal A},
\label{NSR}
\end{eqnarray}
where 
$D_\alpha$ is the covariant derivative for the local supersymmetry and
the index $\alpha$ denote the chiralities of two dimensional spinor. 
$E$ is the superdeterminant of the supervierbein
\cite{pvan}
and it is a scalar density in the superspace.
The supervierbein has been found in \cite{Howe2}:
\begin{eqnarray}
E_m{}^a&=&e_m{}^a+i\bar{\theta}\gamma^a\chi_m+\frac{1}{4}\bar{\theta}\theta{e}_m{}^aA,
\nonumber\\
E_m{}^\alpha&=&\frac{1}{2}\chi_m{}^\alpha+\frac{1}{2}\theta^\mu(\gamma_5)_\mu{}^\alpha\omega_m
-\frac{1}{4}\theta^\mu(\gamma_m)_\mu{}^\alpha{A}-\frac{3i}{16}\bar{\theta}\theta\chi_m{}^\alpha{A}
-\frac{1}{4}\bar{\theta}\theta(\gamma_m)^{\alpha\beta}\phi_\beta,
\nonumber\\
E_\mu{}^a&=&i\theta^\lambda(\gamma^a)_{\lambda\mu},
\nonumber\\
E_\mu{}^\alpha&=&\delta_\mu{}^\alpha-\frac{i}{8}\bar{\theta}\theta\delta_\mu{}^\alpha{A},
\end{eqnarray}
where $e_m{}^a$ is the vierbein and $\chi_m{}^{\alpha}$ 
is the Rarita-Schwinger field
\footnote{Here
one does not need explicit expressions, 
for $\omega_m$ and $\phi^{\beta}$ which are defined in \cite{Howe2},
in which $\phi^{\beta}$ is expressed by $\psi^{\beta}$.
}.
The supergauge transformation of the Rarita-Schwinger field 
contains an auxiliary field, $A$ \cite{Howe2}. 

The invariance of (\ref{NSR}) means $D_\alpha{Y^{\cal B}}D^{\alpha}Y_{\cal B}$ is a scalar in the superspace. 
By integrating the super co-ordinates, we have the action of $NSR$ superstring \cite{GSW}.

\medskip

Next let us consider the following action by keeping the gross structure of (\ref{S_schild}) and supersymmetrizing it,
\begin{eqnarray}
S_{2}=\int{d^2x}{d^2\theta}\{\frac{\Gamma}{E}{\det{h}_{\alpha\beta}}~+~\Delta{E}\},
\label{S2}
\end{eqnarray}
where ${h}_{\alpha\beta}$ is given by 
\begin{eqnarray}
{h}_{\alpha\beta}=D_{\alpha}Y^{\cal C}D_{\beta}Y_{\cal C}.
\end{eqnarray}
The determinant of $h_{\alpha\beta}$ is represented by a bracket:
\begin{eqnarray}
\det{h_{\alpha\beta}}=-\frac{1}{2}\left[Y^{\cal A}~,~Y^{\cal B}\right]^2,
\label{deth}
\end{eqnarray}
where
\begin{eqnarray}
\left[~Y^{\cal A}~,~Y^{\cal B}~\right]=\epsilon^{\alpha\beta}D_{\alpha}Y^{\cal A}D_{\beta}Y^{\cal B}
=D_{\alpha}Y^{\cal A}D^{\alpha}Y^{\cal B}.
\end{eqnarray}
Note that spinor indices, $\alpha$ and $\beta$, are raised and lowered by $\epsilon^{\alpha\beta}$.
The $\det{h}_{\alpha\beta}$ is a scalar in the superspace, 
so that
local symmetries of (\ref{S2}) are restricted to a volume-preserving
diffeomorphism-like transformation in the superspace, in which the transformation parameters $\xi^M$
satisfies
\begin{eqnarray}
\partial_M{\xi}^M=0.
\label{superarea}
\end{eqnarray}
Generally speaking, for a scalar density $G$ transforming as 
$\delta{G}=\partial_n(\epsilon^n{G})$, and
a scalar $I$ transforming as 
$\delta{I}=\epsilon^n\partial_n{I}$, $I^2/G$ behaves under $Diff_2$ as
\begin{eqnarray}
\delta(\frac{I^2}{G})=-\frac{2}{G}\partial_n\epsilon^nI^2+\partial_n(\epsilon^n\frac{I^2}{G}).
\end{eqnarray}
Thus if the infinitesimal group parameter $\epsilon^n$ satisfies $\partial_n\epsilon^n=0$,
then $\delta\{\int{d^2x}I^2/G\}$ becomes a surface term.
In the above example (\ref{S2}), taking $G=E$ and $I=\left[~Y^{\cal A}~,~Y^{\cal B}~\right]$,
we can understand (\ref{superarea}).
The parameters $\xi^M$ have the following expansions \cite{Howe2}:
\begin{eqnarray}
\xi^m&=&f^m-i\bar{\zeta}\gamma^m\theta+\frac{1}{4}\bar{\theta}\theta\bar{\zeta}\gamma^n\gamma^m\chi_n,
\nonumber\\
\xi^\mu&=&\zeta^\mu-\frac{i}{2}\bar{\theta}\gamma^m\zeta\chi_m{}^\mu
-\frac{1}{2}\theta^\lambda(\gamma_5)_\lambda{}^\mu{l}
-\frac{i}{4}\bar{\theta}\theta(\gamma_5\gamma^m)^{\mu\beta}\zeta_\beta\omega_m
-\frac{1}{8}\bar{\theta}\theta\bar{\zeta}\gamma^n\gamma^m\chi_n\chi_m{}^\mu,\nonumber\\
L&=&l-\frac{i}{2}\bar{\zeta}\gamma_5\theta{A}-i\bar{\zeta}\gamma^m\theta\omega_m
+\frac{i}{4}\bar{\theta}\theta\bar{\zeta}\gamma^m\gamma_5\chi_m{A}+\frac{1}{4}\bar{\theta}\theta
\bar{\zeta}\gamma^n\gamma^m\chi_n\omega_m,
\end{eqnarray}
where parameters, $f^m,\zeta^\mu$ and $l$ 
in the first terms correspond to those of $Diff_2$, 
local supersymmetry transformations 
and local Lorentz transformations, respectively.
Notice that (\ref{superarea}) implies four conditions on the five independent parameters, 
$f^m,\zeta^\mu$ and $l$.
By specifically expanding (\ref{superarea}) in terms of $\theta$, we find
\begin{eqnarray}
\partial_m{f}^m+\frac{i}{2}\bar{\chi}_m\gamma^m\zeta&=&0,
\nonumber\\
i\partial_m(\gamma^m\zeta)_\alpha+\frac{i}{2}(\gamma_5\gamma^m\zeta)_\alpha\omega_m
+\frac{1}{4}\chi_{m\alpha}\bar{\zeta}\gamma^n\gamma^m\chi_n&=&0,
\nonumber\\
\partial_m(\bar{\zeta}\gamma^n\gamma^m\chi_n)&=&0.
\label{parameter}
\end{eqnarray}
Terms containing the supergravity multiplet obstruct the locality of solutions for (\ref{parameter}), so that
all gauge degrees of freedom of $Diff_2$ and local supersymmetry
except local Lorentz symmetry have been that can not be moved freely. Only 
gauge parameter $l$ of local Lorentz transformation, is still movable since $\gamma^5$
matrix is trace-less, namely
$(\gamma^5)_\mu{}^\mu=0$.
By integrating the super co-ordinates, we obtain
\begin{eqnarray}
S_{2}=
\int{d}^2x
\frac{\Gamma}{e}[
B_{\cal A}B^{\cal B}\bar{\psi}_{\cal B}\psi^{\cal A}-B^2\bar{\psi}\psi
+\frac{3i}{2}B^{\cal A}\bar{\psi}\psi\bar{\chi}_m
\gamma^m\psi_{\cal A}-2B^{\cal A}\bar{\psi}_{\cal A}\gamma^n\psi^{\cal B}
\partial_n{X}_{\cal B}&&\nonumber\\
+(\bar{\psi}\psi)^2(\frac{3i}{8}A+\frac{1}{2}g^{nm}\bar{\chi_n}{\chi^m}+
\frac{1}{16e}\epsilon^{nm}\bar{\chi_m}\gamma^5\chi_n)
&&\nonumber
\\
+\bar{\psi}\psi(i\bar{\psi}^{\cal B}\gamma^m\partial_m\psi_{\cal B}+g^{nm}
\partial_n{X}_{\cal A}\partial_m{X}^{\cal A}
-\frac{5i}{2}g^{mn}\bar{\chi}_n\psi^{\cal A}\partial_m{X}_{\cal A}
+\frac{i}{2e}\epsilon^{ml}\partial_mX_{\cal A}\bar{\psi^{\cal A}}\gamma^5\chi_l)
&&\nonumber
\\
-\bar{\psi}^{\cal A}\psi^{\cal B}g^{mn}
\partial_m{X}_{\cal A}\partial_n{X}_{\cal B}+\frac{1}{e}\{X_{\cal A},X^{\cal B}\}
\bar{\psi}_{\cal A}\gamma_5\psi^{\cal B}]&&
\nonumber\\
-\int{d}^2x\frac{\Delta}{2}
(\frac{1}{2}\epsilon^{mn}\bar{\chi_m}\gamma_5\chi_n+ieA).
\label{S_2}
\end{eqnarray}
The auxiliary field $A$ appears in (\ref{S_2}) 
because the superconformal symmetry defined by 
Weyl symmetry plus fermionic symmetry has been perfectly restrained.
The quadratic terms of the fermionic matter can be removed from (\ref{S_2}) by using
\begin{eqnarray}
\frac{\delta{S_2}}{\delta{A}}=\frac{3i\Gamma}{8e}
(\bar{\psi}{\psi})^2-\frac{\Delta{i}}{2}e=0.
\end{eqnarray}
The emergence of the Yukawa type interaction, for instance $\epsilon^{mn}\bar{\chi_m}\gamma^5{\chi_n}$,
is one of the interesting features in (\ref{S_2}). Remember that
when one performes a dimensional reduction of the three dimensional $SYM$ theory, 
one finds such a Yukawa type interaction in the dimensional reduced lower dimensions.
It is the remnant of supersymmetry in the higer dimensions.

\bigskip

The intuitive extension of the Schild string (\ref{S_schild}) may be 
\begin{eqnarray}
S_3=\int{d^2x}{d^2\theta}\left[\frac{\Gamma}{E}{sdet}~{h_{MN}}+\Delta{E}\right],
\label{S_3}
\end{eqnarray}
rather than (\ref{S2}).
One may define the induced metric of the superfield propagating in the superspace as 
\begin{eqnarray}
h_{MN}=\partial_M{Y}^{\cal C}\partial_N{Y}_{\cal C}.
\end{eqnarray}
The action (\ref{S_3}) possesses, however, a
pathological behavior because of the superdeterminant \cite{pvan}.
It becomes a non-local functional as,
\begin{eqnarray}
S_3=\int{d^2x}\left[\frac{1}{\{\bar{\psi}\psi\}^2}...............\right].
\label{non-local}
\end{eqnarray}
Therefore a something new device, for example the condensation
of fermion field $\langle\bar{\psi}\psi\rangle\ne{0}$ 
seems to be introduced in order to extract an interesting result from 
(\ref{S_3}).

\section{Summary and discussions}
\indent

We have presented three kinds of deformations 
for the Schild string in the virtue of $N=1$ two dimensional superspace.
The first (\ref{S1}) which is invariant under $N=1$ supersymmetry transformation 
and the area-preserving diffeomorphism seems to include the dynamics of the  
Schild string as we may associate (\ref{boson}) with this.
The second (\ref{S_2}) contains the Yukawa type interaction which reminds us 
two dimensional $SYM$ theory reduced from three dimensions.
The last (\ref{non-local}) which is a non-local functional is formed by the induced metric of superfield. 

\medskip

Next we remark on the difference between the Schild gauge $GS$ superstring action \cite{KKT} and
deformed actions in this letter. 
First remember that
in \cite{Nambu3}
Nambu treated $\{~X_{\cal A}~,~X_{\cal B}~\}^2$ as the square of 
the gauge field strength, $F_{{\cal A}{\cal B}}^2$ under some restrictions.
If one can consistently delineate new Schild action of the $NSR$ superstring,
then the action would look like one of $SYM$ theories 
\footnote{ One of authors (R.K.) would like to thank 
Prof.Y.Kitazawa for pointing out this.}. 
The action of the Schild gauge $GS$ superstring \cite{KKT}
is very similar to the rough structure
of $SYM$ action
\footnote{
In \cite{KKT}
the authors studied not only the kinemaical properties of this but also the dynamical one.
They showed a matrix model like this can be regarded as the continuum limit of the large-$N$ reduced model of
$10$-dimensional $SYM$ theory.}:  
\begin{eqnarray}
S_{SY}=\int{d^Dx}(-\frac{1}{4}F^2+\frac{i}{2}\bar{\psi}{\cal D}\psi),
\label{SY}
\end{eqnarray}
where $\psi$ is the representation of Clifford algebra in the $D$ space-time dimensions.
To hold supersymmetries of (\ref{SY}), spatial-temporal dimension must be only $D=3,4,6$ and $10$ \cite{GSW}.
In order to 
keep the two-dimensional supersymmetry, one needs pseudo scalar and the Yukawa interaction in the action
as we have described in the introduction. 
Therefore it is suitable conclusions that (\ref{S1}), (\ref{S_2}) and (\ref{S_3}) 
are not reduced to the original Schild string 
by usual bosonic truncation. Completing the two dimensional $SYM$ type action by superstring fields 
may not be straightforward as compared with the $GS$ string case \cite{KKT}.

Next we would like to comment on (\ref{S_2}).
Some terms in (\ref{S_2}) may be viewed as the Yukawa coupling with the bound states 
$(\bar{\psi}\psi)^2$ which are formed under the condensation of the fermionic fields.
Let us take notice of the symmetries of (\ref{S_2}).
Its gauge symmetries have been completely fixed except for local Lorentz symmetry.
We would like to improve the whole violation of local symmetries
which reside in the $NSR$ superstring
to a partial break.
The partial break of the symmetries 
is rather desirable disposition.
Because if one can complete a new Schild type action of the $NSR$ superstring,
then it should correspond to a `Nambu-Goto superstring'.
Remember that in the Nambu-Goto string the Weyl symmetry is not surviving
though it is alive  
in the Polyakov string. Therefore to preserve the supersymmetry of the theory, 
the fermionic symmery of $\chi_m~\rightarrow~\chi_m+\gamma_m\eta$ with $\eta$ an arbitrary 
Majorana spinor and the supersymmetry
should be partially restricted. 

Ultimately we close this letter with the following remarks : {\it what are impediments in constructing
new Schild type superstring action from the} $NSR$ {\it superstring action with
vector spinor fields $\psi^{\cal B}_{\mu}$ ?}

\begin{itemize}
\item
The kinematical properties of the Nambu-Goto string is equivalent to the Schild string (\ref{S_schild})
up to the equation of motion
for the auxiliary field $e$. 
Remember that the Nambu-Goto string is derived from the Polyakov string by using the equation of motion
for metric fileds $g_{mn}$ which are not dynamical variables at the classical level.
We can expect that a kind of Schild superstring 
which we have tried to find in this letter has some relation with 
a kind of Nambu-Goto superstring which would be made from the $NSR$ superstring up to equations of
motion for some auxiliary fields, as the Schild string.
So far we have known, such a superstring action like the Nambu-Goto string, that is to say, a kind of 
Nambu-Goto superstring 
with the $D$-plet of Majorana fermions $\psi^{\cal B}_{\mu}$ is not yet completed.
The hardship seems to be as follows
\footnote{One of the authors (R.K.) is inclined to the private communications 
several years ago with Prof. N.Ishibashi \cite{NI} on the relation between $NSR$ superstring and 
supersymmetrized Nambu-Goto string.}.
The action of the $NSR$ superstring contains zwiebein fields $e_m{}^a$ and 
Rarita-Schwinger fields $\chi_{n\alpha}$ as non-dynamical fields at the classical level. 
When one tries to find such a Nambu-Goto superstring,
one should eliminate the non-dynamical supergravity multiplet by using their equations of motion.
One can easily derive these which are, however, non-linear equations. Hence to find their local 
solutions are not generally trivial.
To search some gauge fixing condition in which the equations become linear would be a helpful subject.
For example if we choose $e_m{}^a=\delta^a_m$ on the equation of motion for $\chi_m$
\footnote{
Here we would like to give just a rough discussion to intuitively show the non-linearity.}
,
then the Rarita-Schwinger fields are represented by the matter fields as
\begin{eqnarray}
\chi_m=4i\psi^{\cal A}\partial_mX_{\cal A}/{\bar{\psi}^{\cal B}\psi_{\cal B}}.
\end{eqnarray}
In the derivation, we have used the equation of motion for $\chi_l$,
\begin{eqnarray}
-i\gamma^m\gamma^l\psi^{\cal B}\partial_m{X}_{\cal B}+\frac{1}{4}
\bar{\psi}^{\cal B}\psi_{\cal B}\gamma^n\gamma^l\chi_n=0.
\end{eqnarray}
As one recognizes from this simple case, it is possible that such non-local functions 
will appear in the expression
of a Nambu-Goto superstring. The non-local functions possibly correspond to 
that in the third deformation (\ref{non-local}).

\item
As we have described in the third paragraph of this section,  
fermionic symmetires in the $NSR$ superstring are partially broken in the Schild superstring.
The breaking of the symmetries may be performed by some gauge fixing procedure
in which the fields of $NSR$ superstring are consistently related with the adjoint scalar as well as other
Yang-Mills multiplet, as we have written in the introduction.
Futhermore the gauge fixing conditions should represent linear equations of motion of supergravity multiplet
as we have expressed in the above. 

\item
Creating some kind of Schild superstring with space-time spinors may not be troublesome.
One finds the super version of the constituent $\partial_m{X}^{\cal A}$ of the induced metric,
$\partial_mX^{\cal A}\partial_nX_{\cal A}$, as
\begin{eqnarray}
\Pi^{\cal A}_m\equiv\partial_m{X}^{\cal A}-\bar{\theta}_{\bar{A}}\Gamma^{\cal A}
\partial_m\theta^{\bar{A}},
\label{superpi}
\end{eqnarray}
where $\bar{A}=1,2,...N$ and $\theta$ refers to a space-time spinor \cite{GSW}. 
Expression (\ref{superpi}) is invariant under $N$ global space-time supersymmetires.
For the vector spinor $\psi^{\cal C}_\mu$, 
one can not invent a world-sheet supersymmetric constituent as $\Pi^{\cal A}_m$.
Because it should carry the canonical dimension one,
a space-time index and a world-sheet vector index.
One can not form such a quality by combining the supermultiplet of the $NSR$ superstring 
\footnote{
The vector spinor has
the canonical dimension $1/2$. For example,
such thing $\bar{\psi}^{\cal A}\psi_{\cal B}\partial_m{X}_{\cal A}$ has the canonical dimmensions two.}.
\end{itemize}

\vskip 8mm

{\sl Acknowledgment}

We would like to thank Prof.Y.Kitazawa for reading the origimal manuscript which was titled
`Metamorphoses of Schild String' and his useful discussion.


\end{document}